\documentclass[onecolumn,showpacs,preprintnumbers]{revtex4}

\usepackage{graphics}
\usepackage{epsfig}

\begin{document}

\title{Dilepton radiation by vector mesons and off-shell partons \\ in the hot and dense medium}

\author{O. Linnyk$^a$\thanks{linnyk@th.physik.uni-frankfurt.de}, E. L. Bratkovskaya$^a$
and W. Cassing$^b$ \\
  $^a$Institut f\"ur Theoretische Physik, Universit\"{a}t Frankfurt, 60438 Frankfurt am Main,
  Germany \\
  $^b$Institut f\"ur Theoretische Physik, Universit\"at Giessen,
  35392 Giessen, Germany
}

\begin{abstract}
According to the dynamical quasiparticle model (DQPM) -- matched to
reproduce lattice QCD results in thermodynamic limit, -- the
constituents of the strongly interacting quark-gluon plasma (sQGP)
are massive and off-shell quasi-particles (quarks and gluons) with
broad spectral functions. In order to address the electromagnetic
radiation of the sQGP, we derive off-shell cross sections of $q\bar
q\to\gamma^*$, $q\bar q\to\gamma^*+g$ and $qg\to\gamma^*q$($\bar q
g\to\gamma^* \bar q$) reactions taking into account the effective
propagators for quarks and gluons from the DQPM. Dilepton production
in In+In collisions at 158~AGeV is studied by implementing these
processes into the parton-hadron-string dynamics (PHSD) transport
approach. The microscopic PHSD transport approach describes the full
evolution of the heavy-ion collision: from the dynamics of
quasi-particles in the sQGP phase (when the local energy density is
above $\sim 1$~GeV/fm$^3$) through hadronization and to the
following hadron interactions and off-shell propagation after the
hadronization. A comparison to the data of the NA60 Collaboration
shows that the low mass dilepton spectra are well described by
including a collisional broadening of vector mesons, while the
spectra in the intermediate mass range are dominated by off-shell
quark-antiquark annihilation, quark Bremsstrahlung and gluon-Compton
scattering in the nonperturbative QGP. In particular, the observed
softening of the $m_T$ spectra at intermediate masses (1~GeV~$\le M
\le$~3~GeV) is approximately reproduced.
\end{abstract}

\pacs{ %
25.75.Nq, 
24.10.Lx, 
25.75.Cj,  
14.40.Be,  
24.85.+p   
}


\maketitle


\section{Introduction}

While the  properties of hadrons are rather well known in free space
(embedded in a nonperturbative QCD vacuum) the masses and lifetimes
of hadrons in a baryonic and/or mesonic environment are subject of
current research in order to achieve a better understanding of the
strong interaction and the nature of confinement. A  modification of
vector mesons has been seen experimentally in the enhanced
production of lepton pairs above known sources in nucleus-nucleus
collisions at Super-Proton-Synchroton (SPS) energies
\cite{CERES,HELIOS}. This can be attributed to a shortening of the
lifetime of the vector mesons $\rho$, $\omega$ and $\phi$. The
question arises if the enhancement might (in part) be due to new
radiative channels~\cite{Linnyk:2009nx} from the strong Quark-Gluon
Plasma (sQGP).  The answer to this question is nontrivial due to the
nonequilibrium nature of the heavy-ion reactions and covariant
transport models have to be incorporated to disentangle the various
sources that contribute to the final dilepton spectra seen
experimentally.

\section{The PHSD approach}

To address the vector meson properties in a hot and dense medium --
as created in heavy-ion collisions -- we employ an up-to-date
relativistic transport model, i.e. the Parton Hadron String
Dynamics~\cite{CasBrat} (PHSD) that incorporates the relevant
off-shell dynamics of the vector mesons as well as the explicit
partonic phase in the early hot and dense reaction region. PHSD
consistently describes the full evolution of a relativistic
heavy-ion collision from the initial hard scatterings and string
formation through the dynamical deconfinement phase transition to
the quark-gluon plasma (QGP) as well as hadronization and to the
subsequent interactions in the hadronic phase.

In the hadronic sector PHSD is equivalent to the
Hadron-String-Dynamics (HSD) transport approach
\cite{CBRep98,Brat97,Ehehalt} that has been used for the description
of $pA$ and $AA$ collisions from SIS to RHIC energies and has lead
to a fair reproduction of hadron abundances, rapidity distributions
and transverse momentum spectra. In particular, HSD incorporates
off-shell dynamics for vector mesons -- according to
Refs.~\cite{Cass_off1} -- and a set of vector-meson spectral
functions~\cite{Brat08} that covers possible scenarios for their
in-medium modification. The transport theoretical description of
quarks and gluons in PHSD is based on a dynamical quasiparticle
model for partons matched to reproduce lattice QCD results in
thermodynamic equilibrium (DQPM). The transition from partonic to
hadronic degrees of freedom is described by covariant transition
rates for the fusion of quark-antiquark pairs to mesonic resonanses
or three quarks (antiquarks) to baryonic states.

Various models predict that hadrons change in the (hot and dense)
nuclear medium; in particular, a broadening of the spectral function
or a mass shift of the vector mesons has been expected. Furthermore,
QCD sum rules indicated that a mass shift may lead to a broadening
and vice versa~\cite{MuellerSumRules}; therefore  both modifications
should be studied simultaneously. Thus we explore three possible
scenarios:
 (1) a broadening of
the $\rho$ spectral function, (2) a mass shift, and (3) a broadening
plus a mass shift. The HSD (PHSD) off-shell transport approach
allows to investigate in a consistent way the different scenarios
for the modification of vector mesons in a hot and dense medium. In
the off-shell transport, hadron spectral functions change
dynamically during the propagation through the medium and evolve
towards the on-shell spectral function in the vacuum.

As demonstrated in Ref.~\cite{Brat08} the off-shell dynamics is
particularly important for resonances with a rather long lifetime in
vacuum but strongly decreasing lifetime in the nuclear medium
(especially $\omega$ and $\phi$ mesons) but also proves vital for
the correct description of dilepton decays  of $\rho$ mesons with
masses close to the two pion decay threshold. For a detailed
description of the various hadronic channels included for dilepton
production as well as the off-shell dynamics we refer the reader to
Refs.~\cite{Brat08,here}.

\begin{figure}
\includegraphics[width=0.9\textwidth]{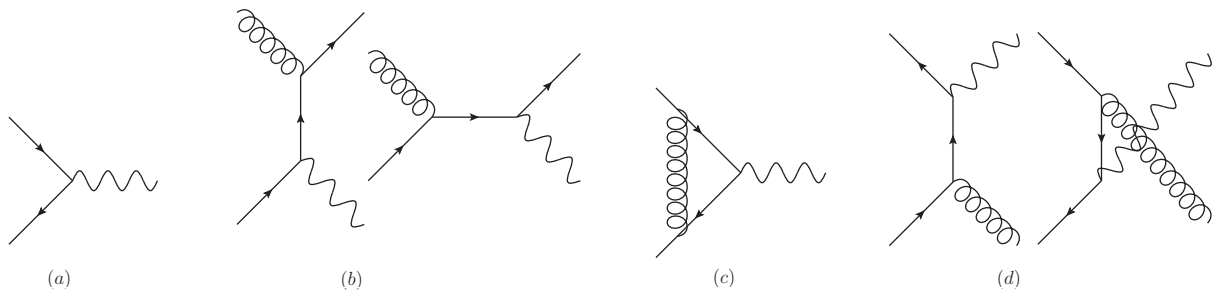}
\caption{Diagrams contributing to the dilepton production from a
QGP: (a) Drell-Yan mechanism, (b) gluon-Compton scattering (GCS),
(c) vertex correction, (d) gluon Bremsstrahlung (NLODY), where
virtual photons (wavy lines) split into lepton pairs, spiral lines
denote gluons, arrows denote quarks. In each diagram the time runs
from left to right.} \label{diagrams}
\end{figure}

\section{Dilepton radiation from the quark gluon plasma}

By employing the HSD approach to the low mass dilepton production in
relativistic heavy-ion collisions, it was shown in~\cite{here} that
the NA60 Collaboration data for the invariant mass spectra for
$\mu^+\mu^-$ pairs from In+In collisions at 158 A$\cdot$GeV favored
the 'melting $\rho$' scenario \cite{NA60}. Also the data from the
CERES Collaboration \cite{CERES2} showed a preference for the
'melting $\rho$' picture. On the other hand, the dilepton spectrum
from In+In collisions at 158 A$\cdot$GeV for $M>1$~GeV could not be
accounted for by the known hadronic sources (see Fig.2
of~\cite{here}). Also, hadronic models do not reproduce the
softening of the $m_T$ distribution of dileptons at
$M>1$~GeV~\cite{NA60}. This observation points towards a partonic
origin but needs further examination.

The transport theoretical description of the partonic dynamics in
PHSD is based on the Dynamical QuasiParticle Model (DQPM)
\cite{Cassing06} which describes QCD properties in terms of
single-particle Green's functions (in the sense of a two-particle
irreducible approach) and leads to effective strongly interacting
partonic quasiparticles with broad spectral functions as degrees of
freedom. Dilepton radiation by the constituents of the strongly
interacting QGP, i.e. by the dynamical quasiparticles, proceeds via
the elementary processes illustrated in Fig.~\ref{diagrams}: the
basic Drell-Yan $q+\bar q$ annihilation mechanism, Gluon Compton
scattering ($q+g\to \gamma^*+q$ and $\bar q+g\to \gamma^*+\bar q$),
and quark + anti-quark annihilation with gluon Bremsstrahlung in the
final state ($q+\bar q\to g+\gamma^*$). Note that in our
calculations the running coupling $\alpha_S$ depends on the local
energy density $\epsilon$ according to the DPQM~\cite{Cassing06},
while the energy density is related to the temperature $T$ by the
lQCD equation of state. Numerically, we observe that $\alpha_S$ is
of the order $O(1)$ and thus the contribution of the higher order
diagrams in Fig.~\ref{diagrams} is not subleading!

In the on-shell approximation, one uses perturbative QCD cross
sections for the processes listed above. However, to make a
quantitative comparison to experimental data at the realistically
low photon virtuality $M\sim 1$~GeV, one has to take into account
the non-perturbative spectral functions and self-energies of the
quarks, anti-quark and gluons thus going beyond the leading twist.
Therefore, we have derived the off-shell cross sections of $q\bar
q\to\gamma^*$, $q\bar q\to\gamma^*+g$ and
$qg\to\gamma^*q$($qg\to\gamma^*q$) reactions by calculating the
corresponding diagrams for massive quarks and gluons, with the
masses being distributed according to the DQPM self energies. The
obtained off-shell elementary cross sections then have been
implemented into the PHSD transport code.
\begin{figure}
  \begin{minipage}[b]{0.495\textwidth}
    \includegraphics[width=\textwidth]{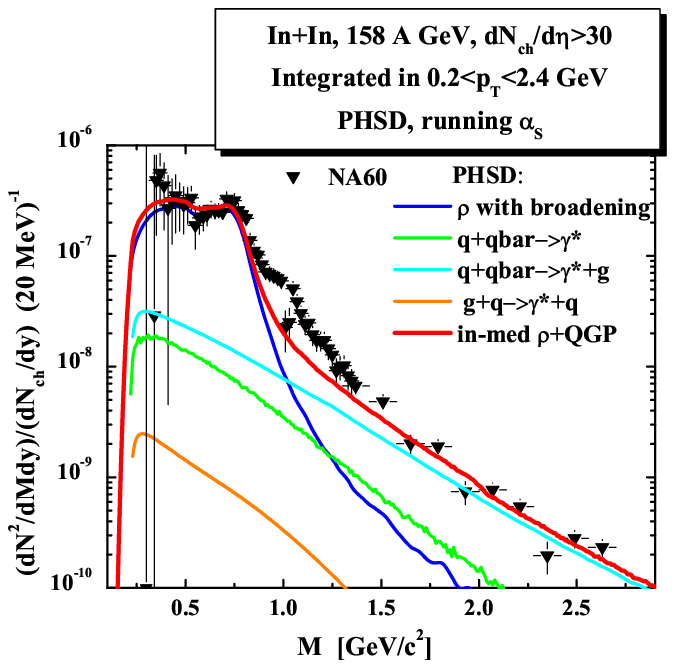}
  \end{minipage}
  \begin{minipage}[b]{0.495\textwidth}
    \includegraphics[width=\textwidth]{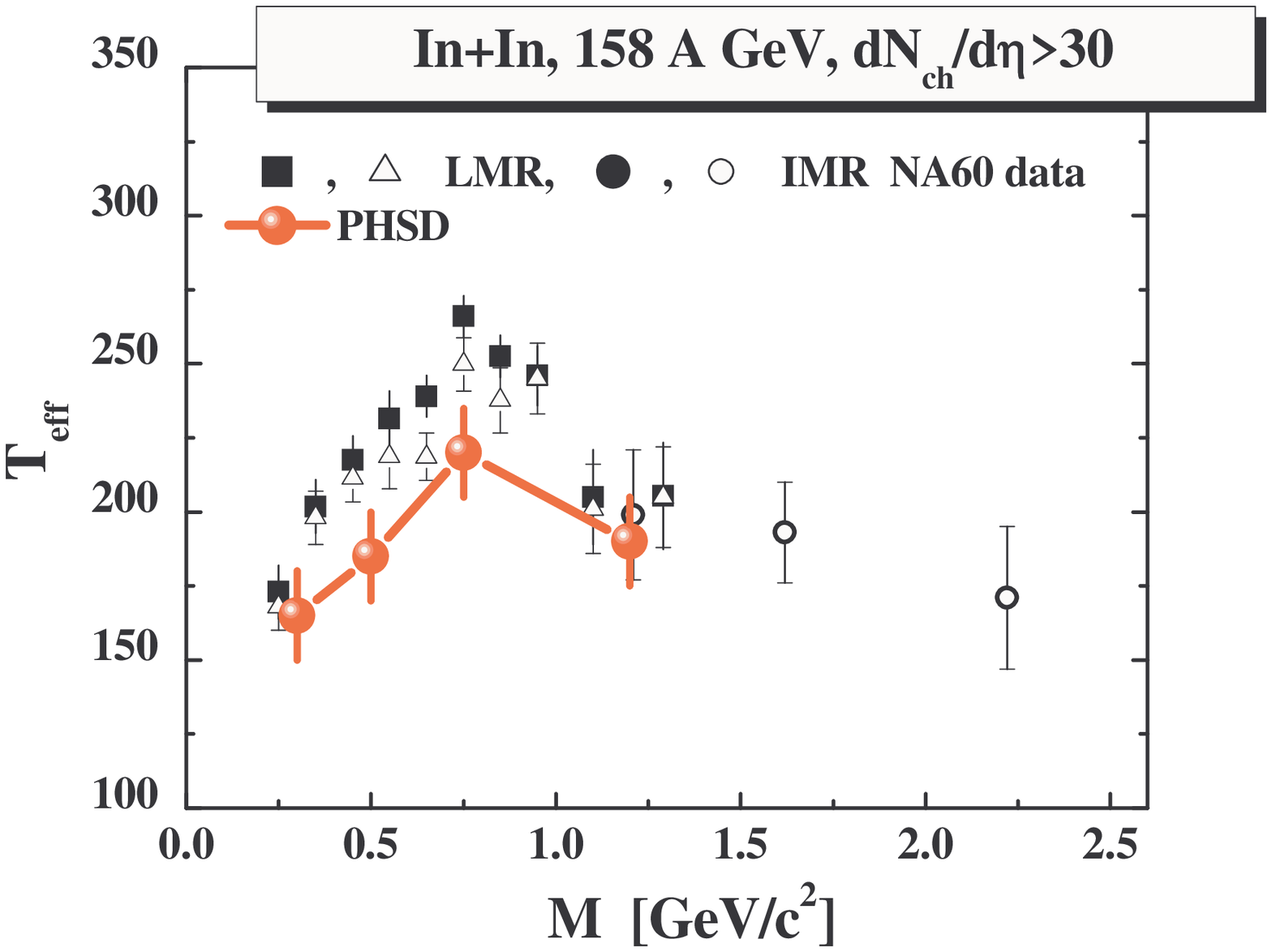}
  \end{minipage}
   \caption{{\bf Left:} Acceptance corrected mass spectra of the excess dimuons from $In+In$
at 158~AGeV integrated over $p_T$ in $0.2<p_T<2.4$~GeV from PHSD
compared to the data of NA60~\cite{Arnaldi:2008er}. {\bf Right:} The
inverse slope parameter $T_{eff}$ of the dimuon yield from In+In at
158 A$\cdot$GeV as a function of the dimuon invariant mass in PHSD
compared to the data of the NA60
Collaboration~\protect{\cite{NA60,Arnaldi:2008er}}. The contribution
from charm decays has been subtracted from the data. }
\label{NA60_AC}
\end{figure}

In Fig.~\ref{NA60_AC}~(L.h.s.) we present PHSD results for the
dilepton spectrum excess over the known hadronic sources as produced
in $In+In$ reactions at 158~AGeV compared to the acceptance
corrected data published recently by the NA60
Collaboration~\cite{Arnaldi:2008er}. We find that the spectrum at
invariant masses below 1~GeV is well reproduced by the $\rho$ meson
yield, if a broadening of the meson spectral function in the medium
is assumed~\cite{here}. Thus the result of~\cite{vanHees} is
confirmed by the PHSD calculations. On the other hand, the spectrum
at $M>1$~GeV is shown to be dominated by the partonic sources.

It is also interesting to note that accounting for partonic dilepton
sources allows to reproduce in PHSD (cf. Fig.~\ref{NA60_AC}, rhs)
the intriguing finding of the NA60
Collaboration~\cite{NA60,Arnaldi:2008er} that the effective
temperature of the dileptons (slope parameters) in the intermediate
mass range is lower than the $T_{eff}$ of the dileptons from the
hadronic phase. The softening of the transverse mass spectrum with
growing invariant mass implies that the partonic channels occur
dominantly before the collective radial flow has developed. In total
we still underestimate the slope parameter $T_{eff}$ (cf. rhs in
Fig.~2) which might be due to missing partonic initial state effects
or an underestimation of partonic flow in the initial phase of the
reaction. These issues are presently being addressed and will be
reported in the near future.
\begin{figure}
\includegraphics[width=.5\textwidth]{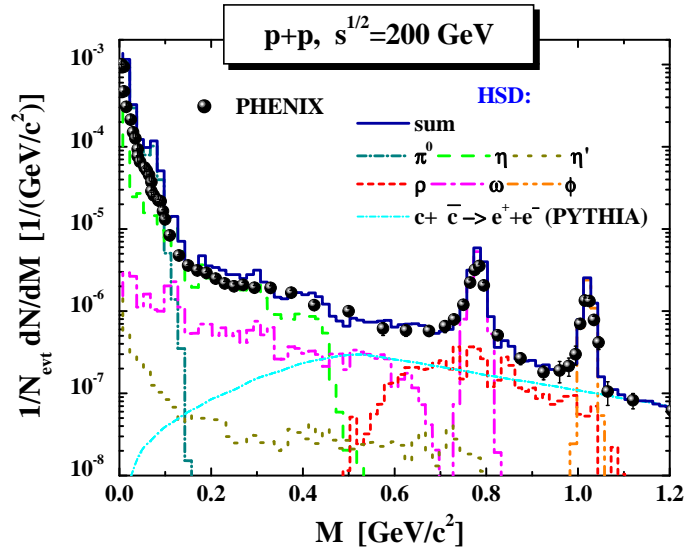}
 \caption{The HSD results  for the mass differential dilepton spectra
in case of $pp$ collisions at $\sqrt{s}$ = 200 GeV  in comparison to
the data from PHENIX~\protect{\cite{PHENIXpp}}. The actual PHENIX
acceptance and mass resolution have been incorporated (see legend
for the different color coding of the individual channels). Figure
taken from~\protect{\cite{here}}.} \label{pp}
\end{figure}

\section{RHIC energies}
\begin{figure}
\includegraphics[height=.6\textheight]{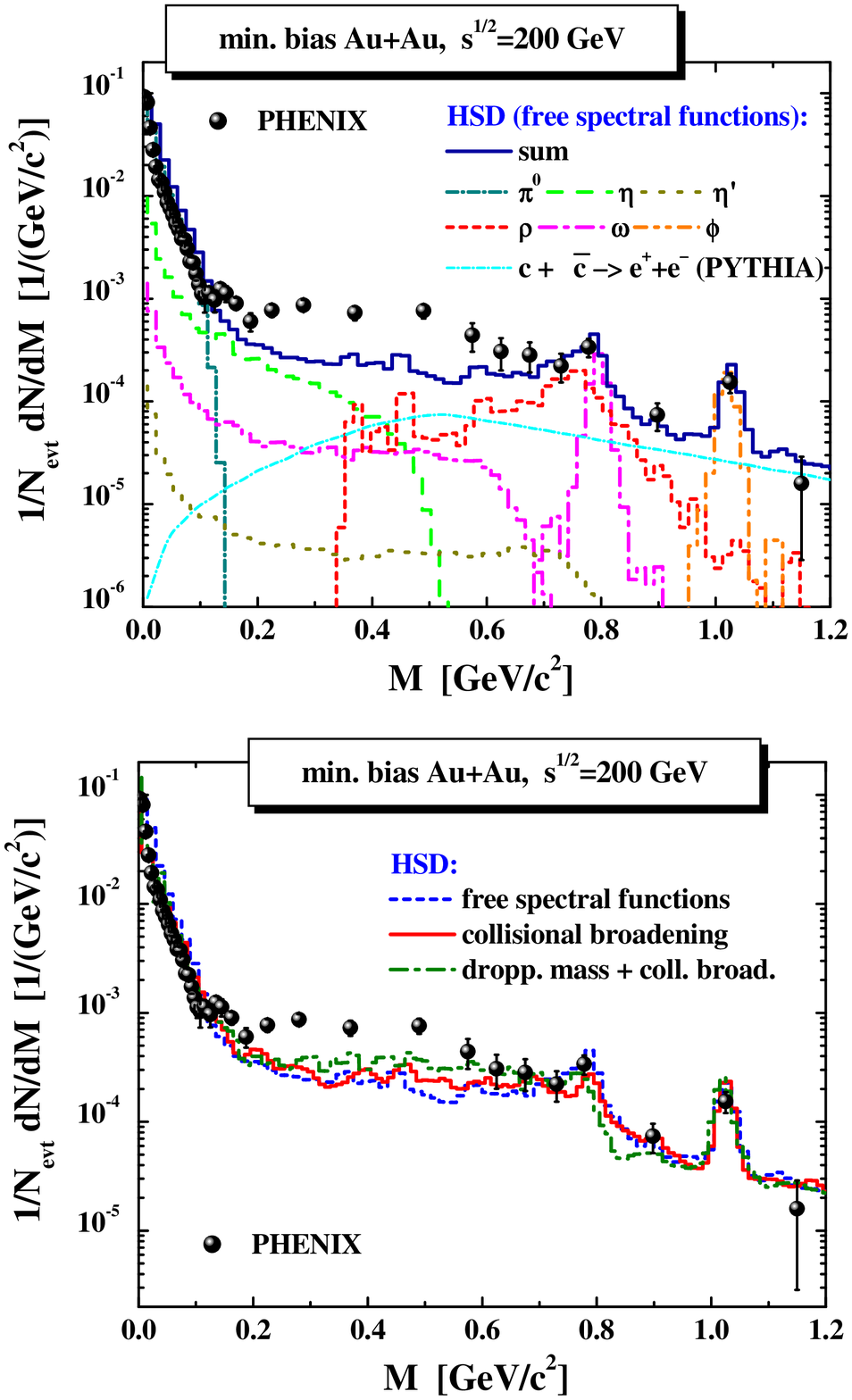}
 \caption{The HSD
results  for the mass differential dilepton spectra in case of
inclusive $Au + Au$ collisions at $\sqrt{s}$ = 200 GeV in comparison
to the data from PHENIX~\protect{\cite{PHENIX}}. The actual PHENIX
acceptance filter and mass resolution have been
incorporated~\protect{\cite{Alberica}}. In the upper part the
results are shown for  vacuum spectral functions (for $\rho, \omega,
\phi$) including the channel decompositions (see legend for the
different color coding of the individual channels). The lower part
shows a comparison for the total $e^+e^-$ mass spectrum in case of
the 'free' scenario (dashed line), the 'collisional broadening'
picture (solid line) as well as the 'dropping mass + collisional
broadening' model (dash-dotted line). Figure taken
from~\protect{\cite{here}}. } \label{Fig3}
\end{figure}

In 2008, the PHENIX Collaboration has presented first dilepton data
from $pp$ and $Au+Au$ collisions at Relativistic-Heavy-Ion-Collider
(RHIC) energies of $\sqrt{s}$ = 200 GeV \cite{PHENIX} which show an
even larger enhancement in $Au+Au$ reactions (relative to $pp$
collisions) in the invariant mass regime from 0.15 to 0.6 GeV than
the data at SPS energies \cite{NA60CERES2}. We recall that HSD
provides a reasonable description of hadron production in $Au+Au$
collisions at $\sqrt{s}$ = 200 GeV~\cite{Brat03} such that we can
directly continue with the results for $e^+e^-$ pairs which are
shown in Fig.~\ref{pp} for $p+p$ collisions. We find that the
dilepton radiation in the elementary channel is well under control
in HSD, which is equivalent to PHSD for $p+p$ reactions.

We step on with the case of inclusive $Au + Au$ collisions in
comparison to the data from PHENIX~\cite{PHENIX} (Fig~\ref{Fig3}).
When including the in-medium modification scenarios for the vector
mesons, we achieve a sum spectrum which is only slightly enhanced
compared to the 'free' scenario. Whereas the total yield  is quite
well described in the region of the pion Dalitz decay as well as
around the $\omega$ and $\phi$ mass, HSD clearly underestimates the
measured spectra in the regime from 0.2 to 0.6 GeV by an average
factor of 3. The low mass dilepton spectra from $Au+Au$ collisions
at RHIC (from the PHENIX Collaboration) are clearly underestimated
in the invariant mass range from 0.2 to 0.6 GeV in the 'collisional
broadening' scenario as well as in the 'dropping mass + collisional
broadening' model, i.e. when assuming a shift of the vector meson
mass poles with the baryon density. We mention that our results for
the low mass dileptons are very close to the calculated spectra from
van Hees and Rapp as well as Dusling and Zahed~\cite{Dussi} (cf. the
comparison in Ref.~\cite{AToia}). Consequently we attribute this
additional low mass enhancement seen by PHENIX to non-hadronic
sources and will proceed with the quantitative test of this
assumption in a future publication.


\section*{Acknowledgments}

OL acknowledges financial support within the ``HIC for FAIR" center
of the ``LOEWE'' program.


\begin{thebibliography}{99}
\bibitem{CERES}
        G. Agakichiev {\it et al.}, CERES Collaboration,
              \emph{ Phys. Rev. Lett.} {\bf 75} (1995) 1272;
        Th. Ullrich {\it et al.},  \emph{ Nucl. Phys.} { A}{\bf 610} (1996) 317c;
        A. Drees, \emph{ Nucl. Phys.} { A}{\bf 610} (1996) 536c.
\bibitem{HELIOS}
        M. A. Mazzoni, HELIOS Collaboration,
              \emph{ Nucl. Phys.} { A}{\bf 566} (1994) 95c;
        M. Masera, \emph{ Nucl. Phys.} { A}{\bf 590} (1995) 93c;
        T. {\AA}kesson et al., \emph{ Z. Phys.} { C}{\bf 68} (1995) 47.
\bibitem{Linnyk:2009nx}
  O.~Linnyk, E.~L.~Bratkovskaya and W.~Cassing,
  \emph{Nucl.\ Phys.}\  A {\bf 830} (2009) 491C.
\bibitem{CasBrat} W. Cassing and E. L. Bratkovskaya,
    \emph{Phys. Rev.} C {\bf 78} (2008) 034919,
  \emph{Nucl.\ Phys.}\  A {\bf 831} (2009) 215.
\bibitem{CBRep98}
        W. Cassing, E. L. Bratkovskaya,
        \emph{ Phys. Rept.} {\bf 308} (1999) 65.
\bibitem{Brat97}
        E. L. Bratkovskaya, W. Cassing,
      \emph{ Nucl. Phys.} { A}{\bf 619} (1997) 413.
\bibitem{Ehehalt}
    W. Ehehalt, W. Cassing, \emph{ Nucl. Phys.} { A }{\bf 602} (1996) 449.
\bibitem{Cass_off1}
    W. Cassing, S. Juchem, \emph{ Nucl. Phys.} { A }{\bf 665} (2000)
    377; {\it ibid.} { A }{\bf 672} (2000) 417.
\bibitem{MuellerSumRules}
  J.~Ruppert, T.~Renk and B.~M{\"u}ller,
  \emph{Phys.\ Rev.}\  C {73} (2006)  034907.
\bibitem{Brat08}
       E. L. Bratkovskaya, W. Cassing,
      \emph{Nucl. Phys.} A {\bf 807} (2008) 214.
\bibitem{here}
  E. L. Bratkovskaya, W. Cassing and O. Linnyk,
  \emph{Phys.\ Lett.}\  B {\bf 670} (2009) 428.
\bibitem{vanHees}
  H.~van Hees and R.~Rapp,
  \emph{Nucl.\ Phys.}\  A {\bf 806} (2008) 339;
  \emph{Phys.\ Rev.\ Lett.}\  {\bf 97} (2006) 102301.
\bibitem{Cassing06}
      W. Cassing, \emph{Nucl. Phys.} A 791 (2007) 365; {\it ibid.}  A 795 (2007) 70.
\bibitem{NA60}
      R. Arnaldi {\it et al.}, NA60 Collaboration,
              \emph{Phys. Rev. Lett.} {\bf 96} (2006)   162302;
      J. Seixas {\it et al.},  \emph{J. Phys.} G {\bf 34} (2007) S1023;
       S. Damjanovic {\it et al.}, \emph{Nucl. Phys.} {\bf A 783} (2007) 327c.
\bibitem{CERES2}
       D. Adamova {\it et al.} CERES Collaboration,
         \emph{Nucl. Phys.} A {\bf 715} (2003) 262;
         \emph{Phys. Rev. Lett.} {\bf 91} (2003) 042301;
       G. Agakichiev {\it et al.},
         \emph{Eur. Phys. J.} C {\bf 41} (2005) 475;
       D. Adamova {\it et al.}
  \emph{Phys.\ Lett.}\  B {\bf 666} (2008) 425;
       A. Marin {\it et al.}; Proceedings of CPOD07,
              PoS 034 (2007).
\bibitem{Arnaldi:2008er} 
  R.~Arnaldi {\it et al.}, NA60 Collaboration,
  Eur.\ Phys.\ J.\  C {\bf 59}, 607 (2009)
\bibitem{PHENIXpp}
       A. Adare {\it et al.}, PHENIX Collaboration,
  \emph{Phys.\ Lett.}\  B {\bf 670} (2009) 313
\bibitem{PHENIX}
       A. Toia {\it et al.},  PHENIX Collaboration,
       \emph{Nucl. Phys.} A {\bf 774} (2006) 743;
       \emph{Eur. Phys.} J {\bf 49} (2007) 243;
      S. Afanasiev {\it et al.}, PHENIX Collaboration,  arXiv:0706.3034 [nucl-ex];
   A.~Adare {\it et al.}, PHENIX Collaboration,
  arXiv:0912.0244 [nucl-ex].
\bibitem{Alberica} A. Toia, private communication.
\bibitem{Brat03}
       E. L. Bratkovskaya, W. Cassing, H. St\"ocker,
       \emph{Phys. Rev. } {\bf C 67}  (2003) 054905;
       E. L. Bratkovskaya {\it et al.}
       \emph{Phys. Rev. } {\bf C 69} (2004) 054907.
\bibitem{Dussi}
  K.~Dusling and I.~Zahed,
  \emph{Nucl.\ Phys.}\  A {\bf 825} (2009) 212.
\bibitem{AToia}
  A.~Toia,
  \emph{J.\ Phys.}\ G {\bf 35} (2008) 104037.
\end{thebibliography}
\end{document}